\documentclass[prb,twocolumn,amsmath,amssymb,floatfix]{revtex4}

\usepackage{rotating} 
\usepackage{bm} 
\usepackage{graphicx} 
\newcommand{\df}{$\Delta F$}

\begin{document}

\title{Comparison of free energy methods for molecular systems}
\author{F.\ Marty Ytreberg\footnote{E-mail: fmytreberg@gmail.com}}
\affiliation{Department of Physics,
    University of Idaho, Moscow, ID 83844-0903}
\author{Robert H.\ Swendsen}
\affiliation{Department of Physics, Carnegie Mellon University,
    Pittsburgh, PA 15213}
\author{Daniel M.\ Zuckerman\footnote{E-mail: dmz@ccbb.pitt.edu}}
\affiliation{Department of Computational Biology,
    University of Pittsburgh, 3064 BST-3, Pittsburgh, PA 15213}
\date{\today}

\begin{abstract}
We present a detailed comparison of computational efficiency and precision
for several free energy difference (\df) methods.
The analysis includes both equilibrium and non-equilibrium approaches,
and distinguishes between uni-directional and bi-directional 
methodologies.
We are primarily interested in comparing two recently proposed approaches,
adaptive integration and single-ensemble path sampling, to more
established methodologies.
As test cases, we study relative solvation
free energies, of large changes to the size or charge
of a Lennard-Jones particle in explicit water.
The results show that, for the systems used in this study, both adaptive
integration and path sampling offer unique advantages over the more
traditional approaches.
Specifically, adaptive integration is found to provide very precise
long-simulation \df\ estimates as compared to other methods used in
this report, while also offering rapid estimation of \df.
The results demonstrate that the adaptive integration approach is
the best overall method for the systems studied here.
The single-ensemble path sampling approach is found to be
superior to ordinary Jarzynski averaging for the 
uni-directional, ``fast-growth'' non-equilibrium case.
Closer examination of the path sampling approach on a two-dimensional
system suggests it may be the overall method of choice when
conformational sampling barriers are high.
However, it appears that the free energy
landscapes for the systems used in this study
have rather modest configurational sampling barriers.
\end{abstract}
\maketitle

\section{Introduction}
Free energy difference (\df) calculations are useful for a wide variety
of applications, including drug design \cite{jorgensen-sci,burgers},
solubility of small molecules \cite{vangunsteren-onestep,grossfield-jacs},
and protein/ligand binding affinities
\cite{kollman-pnas,vangunsteren-estrogen,shirts-binding}.
Due to the high computational cost of \df\ calculations, it is of interest
to carefully compare the efficiencies of the various approaches.

We are particularly interested in assessing recently proposed methods
\cite{swendsen-aim,ytreberg-seps} in comparison to established
techniques. Thus, the purpose of this
study is to provide a careful comparison of the efficiency and
precision of several \df\ methods.
We seek to answer two important questions:
(i) Given a fixed amount of computational
time ($10^6$ dynamics steps, in this study),
which method estimates the correct value of \df\ with the greatest precision?
(ii) Which \df\ approach can obtain a ``reasonable'' estimate of \df\ in
the least amount of computational time?

Free energy difference methods can be classified as either equilibrium or
non-equilibrium. Equilibrium approaches include multi-stage free 
energy perturbation
\cite{zwanzig}, thermodynamic integration \cite{kirkwood,straatsma-ti},
Bennett analysis \cite{bennett,shirts-benn} and
weighted histogram analysis \cite{swendsen-wham}. The common theme
in these approaches is that sufficiently
long equilibrium simulations are
performed at each intermediate stage of the free energy calculation.
Equilibrium methods are in wide
use and are known to provide accurate results; however, the computational 
cost can be large due the simulation time needed to attain equilibrium at 
each intermediate stage.
A host of non-equilibrium methods have recently been applied to various
molecular systems, largely due to Jarzynski's remarkable equality
\cite{jarzynski,crooks-pre}.
Non-equilibrium methods have the potential to provide very rapid estimates 
of \df, but can suffer from significant bias
\cite{ytreberg-extrap,zuckerman-prl,hummer}.

In this report we present results using both equilibrium and non-equilibrium 
approaches---as well as uni-directional and bi-directional methodology.
Specifically, we compare:
(i) adaptive integration \cite{swendsen-aim};
(ii) thermodynamic integration \cite{kirkwood};
(iii) single-ensemble path sampling of non-equilibrium work values
using Jarzynski's uni-directional averaging \cite{ytreberg-seps};
(iv) single-ensemble path sampling using Bennett's bi-directional
formalism;
(v) Jarzynski averaging of non-equilibrium work values
\cite{jarzynski,jarzynski-pre}; 
(vi) Bennett analysis of non-equilibrium work values
\cite{crooks-pre,shirts-prl};
(vii) equilibrium Bennett analysis \cite{bennett,shirts-benn}; and
(viii) multi-stage free energy perturbation \cite{zwanzig}.
We also compare the free energy profiles, which determines
the potential of mean force, for adaptive integration and
thermodynamic integration.

Generally, one is interested in the free energy difference
($\Delta F = F_1 - F_0$) between two states or
systems of interest denoted by potential energy functions
$U_0(\vec{x})$ and $U_1(\vec{x})$, where
$\vec{x}$ is the full set of configurational coordinates. \df\ can
be written in terms of the partition functions for each state
\begin{eqnarray}
    \Delta F = -k_B T \ln \Bigg(
	\frac{Z\big[ U_1(\vec{x}) \big]}{Z\big[ U_0(\vec{x}) \big]}
    \Bigg),
    \label{eq-zratio}
\end{eqnarray}
where $k_B$ is the Boltzmann constant, $T$ is the system temperature,
and $Z[U(\vec{x})] = \int d\vec{x} \exp[-U(\vec{x})/k_BT]$.
Because the overlap between the configurations in $U_0$ and $U_1$ may be
poor, a ``path'' connecting $U_0$ and $U_1$ is typically created.
In our notation, the path will be parameterized using the
variable $\lambda$, with $0 \leq \lambda \leq 1$.

\section{Equilibrium free energy calculation}
Equilibrium free energy methodologies share the common strategy of
generating equilibrium ensembles of configurations at multiple
values of the scaling parameter $\lambda$. In the current study
we investigate thermodynamic integration \cite{kirkwood},
adaptive integration \cite{swendsen-aim},
multi-stage free energy perturbation \cite{zwanzig},
and multi-stage equilibrium Bennett analysis \cite{bennett}.
We performed separate equilibrium simulations at successive
values of $\lambda$,
and then estimated \df\ using free energy perturbation,
Bennett averaging, and thermodynamic integration
on the resulting ensemble of configurations
(detailed in Sec.\ \ref{sec-details}).

\subsection{\label{sec-ti}Thermodynamic integration}
Thermodynamic integration (TI) is probably the most
common fully equilibrium \df\ approach. In TI, equilibrium
simulations are performed at multiple values of $\lambda$.
Then, \df\ is found by approximating
the integral \cite{kirkwood},
\begin{eqnarray}
  \Delta F=
  \int_{\lambda=0}^1 d \lambda \left<
  \frac{\partial U_{\lambda}({\vec{x}})}{\partial \lambda} \right>_{\lambda},
  \label{eq-ti}
\end{eqnarray}
where the functional form for $U_\lambda(\vec{x})$ depends upon the
scaling methodology and will be discussed in detail in Sec.\
\ref{sec-details}. 
The notation $\langle ... \rangle_\lambda$ indicates
an ensemble average at a particular value of $\lambda$.
In addition to the possibility of inadequate equilibrium sampling 
at each $\lambda$ value,
error arises in TI from the fact that only a finite number of $\lambda$ values
can be simulated, and thus the integral must be approximated by a sum 
\cite{shirts-benn}.
Thermodynamic integration can provide very accurate \df\
calculations, but can also be computationally expensive
due to the equilibrium sampling required at each $\lambda$ value
\cite{karplus-jpcB,mccammon-ti,shirts-jcp,lybrand}.

\subsection{\label{sec-aim}Adaptive integration}
The adaptive integration method (AIM), detailed in Ref.\
\onlinecite{swendsen-aim},
seeks to estimate the same integral as that of TI; namely Eq.\ (\ref{eq-ti})
(see also discussions in Refs.\
\onlinecite{marinari,tidor-ldyn,brooks-ldyn,wang,deem-aim}).
However, in addition to fixed-$\lambda$ equilibrium sampling,
the AIM approach uses a Metropolis Monte Carlo procedure to generate
equilibrium ensembles for the set of $\lambda$ values.
The $\lambda$-sampling is
done by attempting Monte Carlo
moves that change the value of $\lambda$
during the simulation. The probability of accepting a
change from the old value $\lambda_o$ to a new value $\lambda_n$ is
\begin{eqnarray}
    P_{\rm acc}(\lambda_o \rightarrow \lambda_n) = \nonumber \\
	\min \left[1.0,
	e^{-\beta \bigl( U_{\lambda_n}(\vec{x})-U_{\lambda_o}(\vec{x}) \bigr)}
	e^{+\beta \bigl( \delta \hat{F}(\lambda_n)-\delta 
	    \hat{F}(\lambda_o) \bigr)}
    \right],
    \label{eq-Paim}
\end{eqnarray}
where $\beta=1/k_B T$ and $\delta \hat{F}(\lambda_i)$ is the
{\it current} running free energy estimate
obtained by numerically approximating the integral
\begin{eqnarray}
    \delta \hat{F} (\lambda_i) =
    \int_{\lambda=0}^i d \lambda \left<
    \frac{\partial U_{\lambda}({\vec{x}})}{\partial \lambda} \right>_{\lambda}.
    \label{eq-dFaim}
\end{eqnarray}
Between attempted Monte Carlo moves in $\lambda$, any canonical sampling
scheme (e.g., molecular dynamics, Langevin dynamics, Monte Carlo) can be
used to propagate the system at fixed $\lambda$.
In this report, Langevin dynamics is used to
sample configurations, and Monte Carlo moves in
$\lambda$ are attempted after every time step.

It is important to note that, due to the use of the running estimate 
$\delta \hat{F}$ in Eq.\ (\ref{eq-Paim}), the AIM method satisfies detailed 
balance only asymptotically. In other words, once the \df\ estimate fully 
converges, the value of $\delta \hat{F}$
is correct, and detailed balance is satisfied \cite{swendsen-aim,deem-aim}.

AIM is related to parallel tempering simulation \cite{marinari},
and has the associated advantage: equilibrium sampling of conformational
space at one $\lambda$ value can assist sampling at other $\lambda$ values
due to the frequent $\lambda$ moves. This is reminiscent of
``$\lambda$ dynamics'' simulation \cite{tidor-ldyn,brooks-ldyn}, but contrasts
with TI where only a single starting configuration is passed between
$\lambda$ values.

An additional advantage of AIM over the other methods detailed in
this report is that there is a simple, built-in, reliable,
convergence criterion. Specifically, one can keep track of the population
(number of simulation snapshots) at each value of $\lambda$.
When the estimate for \df\ has converged,
the population will be approximately
uniform across all values of $\lambda$.
If the population is not approximately uniform,
then the simulation must be continued.

\subsection{\label{sec-fep}Free energy perturbation}
In the free energy perturbation approach, one
performs independent equilibrium
simulations at each $\lambda$ value (like TI), then uses
exponential averaging to determine the free energy difference
between neighboring 
$\lambda$ values \cite{zwanzig}---these differences
are then summed to obtain the total
free energy difference. \df\ can be approximated
for a path containing $n$ $\lambda$-values (including
$\lambda=0$ and $\lambda=1$) using the ``forward'' estimate (FEPF)
\begin{eqnarray}
    \Delta F = -k_BT \sum_{i=0}^{n-1}\ln \Bigl<
	e^{-\beta(U_{\lambda_{i+1}}(\vec{x}_i)-U_{\lambda_i}(\vec{x}_i)}
    \Bigr>_{\lambda_i},
    \label{eq-fepf}
\end{eqnarray}
or the ``reverse'' estimate (FEPR)
\begin{eqnarray}
    \Delta F = +k_BT \sum_{i=0}^{n-1} \ln \Bigl<
	e^{-\beta(U_{\lambda_i}(\vec{x}_{i+1})-
	    U_{\lambda_{i+1}}(\vec{x}_{i+1})}
    \Bigr>_{\lambda_{i+1}}.
    \label{eq-fepr}
\end{eqnarray}
A primary limitation of free energy perturbation is that
the spacing between $\lambda$
values must be small enough that there is sufficient overlap between
all pairs ($\lambda_i,\;\lambda_{i+1}$) of configuration spaces.

\subsection{\label{sec-benn}Equilibrium Bennett estimation}
It is also possible to use Bennett's method to combine the information
normally used for forward and reverse free energy perturbation.
In this approach, one computes the free energy difference
between successive $\lambda$ values $\delta F_i$ according to
\begin{eqnarray}
    \Bigg<\bigg[ 1 + e^{\beta \big(
	U_{\lambda_{i+1}}(\vec{x}_i)-U_{\lambda_i}(\vec{x}_i)
	- \delta F_i \big)}
    \bigg]^{-1}\Bigg>_{\lambda_i} = \nonumber \\
    \Bigg<\bigg[ 1 + e^{\beta \big(
	U_{\lambda_{i+1}}(\vec{x}_{i+1})-U_{\lambda_i}(\vec{x}_{i+1})
	+ \delta F_i \big)}
    \bigg]^{-1}\Bigg>_{\lambda{i+1}}.
\end{eqnarray}
Then the sum of these
$\delta F_i$ is the total free energy difference \cite{bennett};
\begin{eqnarray}
    \Delta F = \sum_{i=0}^{n-1} \delta F_i.
    \label{eq-benn}
\end{eqnarray}
Studies have shown that using the Bennett method to evaluate free
energy data is the most efficient manner to utilize two
equilibrium ensembles \cite{shirts-benn,lu-2003}.

\section{Non-equilibrium free energy estimation}
In non-equilibrium free energy approaches, the system is forced
to switch to subsequent $\lambda$ values, whether
or not equilibrium has been reached at the current
$\lambda$ value. In this way, non-equilibrium paths are generated
that connect $U_0$ and $U_1$.
In the current study we use uni-directional Jarzynski averaging
\cite{jarzynski} and bi-directional Bennett averaging of
Jarzynski-style work values \cite{crooks-pre},
as well as uni-directional \cite{ytreberg-seps}
and bi-directional averaging of path sampled work values.

\subsection{\label{sec-jarz}Jarzynski averaging}
For the Jarzynski method \cite{jarzynski},
one considers non-equilibrium paths that alternate between increments
in $\lambda$ and ``traditional'' dynamics 
(e.g., Monte Carlo or molecular dynamics) in $\vec{x}$ at fixed 
$\lambda$ values. Thus, a path with $n$ $\lambda$-steps is given by
\begin{eqnarray}
  {\bf Z}_n = \Bigl\{
    (\lambda_0=0,\vec{x}_0),(\lambda_1,\vec{x}_0),
    (\lambda_1,\vec{x}_1),(\lambda_2,\vec{x}_1), \nonumber \\
    (\lambda_2,\vec{x}_2),...,
    (\lambda_{n-1},\vec{x}_{n-1}),(\lambda_n=1,\vec{x}_{n-1})
    \Bigr\},
  \label{eq-traj}
\end{eqnarray}
where it should be noted that increments (steps) from
$\lambda_i$ to $\lambda_{i+1}$
are performed at a fixed conformation $\vec{x}_i$, and
the initial $\vec{x}_0$ is drawn from
the canonical $U_0$ distribution. For simplicity, Eq.\ \eqref{eq-traj}
shows only a single dynamics step performed at each fixed $\lambda_i$,
from $\vec{x}_{i-1}$ to $\vec{x}_i$; However, 
multiple steps may be implemented, as below (Sec.\ \ref{sec-results}).
A ``forward'' work value is thus given by
\begin{eqnarray}
  W_{\rm f}({\bf Z}_n)=\sum_{i=0}^{n-1}\Bigl[
    U_{\lambda_{i+1}}(\vec{x}_i)-U_{\lambda_i}(\vec{x}_i)
    \Bigr].
  \label{eq-Wjarz}
\end{eqnarray}

By generating multiple paths (and thus work values) it is
possible to estimate \df\ via Jarzynski's equality \cite{jarzynski}
\begin{eqnarray}
    \Delta F = -k_B T \ln \left< e^{-\beta W_{\rm f}} \right>_0,
    \label{eq-jarz}
\end{eqnarray}
where the $\left< ... \right>_0$ represents an average over forward work
values $W_{\rm f}$ generated by starting the system at $U_0$ and 
ending at $U_1$.
A similar expression can be written for the situation when work values 
are generated by switching from $U_1$ to $U_0$.
This approach is ``uni-directional'' since only work values 
from either forward or reverse data are used.

Perhaps the most remarkable aspect of Eq.\ (\ref{eq-jarz})
is that it is valid for arbitrary switching speed. However, in practice,
the \df\ estimates are very sensitive to the distribution of work values, 
which in turn is largely dependent on the switching speed. If 
the distribution of work values is non-Gaussian and the width is large 
($\sigma_W \gg k_B T$), then the \df\ estimate can be heavily biased
\cite{hummer,bustamante-pnas,zuckerman-prl,ytreberg-extrap}.
Consistent with results in this report (Sec.\ \ref{sec-results}),
other efficiency studies \cite{hummer,crooks-pre} have suggested 
that the optimal efficiency
for uni-directional Jarzynski averaging is when the switching speed is slow
enough that $\sigma_W \approx 1 \; k_B T$.

\subsection{\label{sec-bjarz}Bennett averaging of Jarzynski work values}
Due to the bias introduced in using uni-directional Jarzynski averaging, it is
useful to consider a method where both forward and reverse work values
are utilized.
It has been shown that the most efficient use of bi-directional data is via
Bennett's method \cite{crooks-pre,shirts-prl},
\begin{eqnarray}
    \sum_{N_{\rm f}}
	\bigg[ 1 + e^{\beta \big( \eta + W_{\rm f} - \Delta F \big)}
	\bigg]^{-1} =
    \sum_{N_{\rm r}}
	\bigg[ 1 + e^{\beta \big( -\eta + W_{\rm r} + \Delta F \big)}
	\bigg]^{-1},
    \label{eq-bjarz}
\end{eqnarray}
where $\eta=k_B T \ln \left( N_{\rm f}/N_{\rm r} \right)$ allows for
differing number of forward ($N_{\rm f}$) and reverse 
($N_{\rm r}$) work values.
Equation (\ref{eq-bjarz}) must be solved iteratively since \df\ 
appears in the sum on both sides of the equation.

\subsection{\label{sec-seps}Single-ensemble path sampling}
Single-ensemble path sampling (SEPS) is a non-equilibrium approach that
seeks to generate ``important'' paths more frequently
\cite{sun,sun-2004,athenes-pre,athenes-epj,athenes,ytreberg-seps}.
The method uses importance sampling to generate paths
(and thus work values) according to an arbitrary distribution $D$,
here chosen as \cite{ytreberg-seps}
\begin{eqnarray}
    D({\bf Z}_n) = Q({\bf Z}_n) \; e^{-\frac{1}{2} \beta W({\bf Z}_n)},
    \label{eq-sepsdist}
\end{eqnarray}
where $Q({\bf Z}_n)$ is proportional to the probability of occurrence of an
ordinary Jarzynski path, and is given below.
With this choice of $D$ the free energy is estimated via
(compare to Refs.\ \onlinecite{sun,sun-2004,athenes-pre,athenes-epj,athenes})
\begin{eqnarray}
    \Delta F = -k_B T \ln \bigg[
	{\sum}^D e^{-\frac{1}{2}\beta W_{\rm f}}
	\bigg/ {\sum}^D e^{+\frac{1}{2} \beta W_{\rm f}} \bigg],
    \label{eq-seps}
\end{eqnarray}
where the ${\sum}^D$ is a reminder that the work values used in the
sum must be generated according to the distribution in
Eq.\ (\ref{eq-sepsdist}).
Since forward work values, $W_f$ are utilized in Eq.\ \eqref{eq-seps},
the paths must start in $U_0$ and end in $U_1$.
A similar expression can be written for 
reverse work values $W_{\rm r}$.

To generate work values according to the distribution $D$, path sampling
must be used
\cite{pratt,tps-review,hummer-tps,sun,sun-2004,
athenes-pre,athenes-epj,athenes,ytreberg-seps}.
In path sampling, entire paths are generated and then
accepted or rejected according to a suitable Monte Carlo criteria.
In general, the probability of accepting a trial path
with $n$ $\lambda$-steps
(${\bf Z}_n'$ with work value $W'$) that was generated from an
existing path (${\bf Z}_n$ with work value $W$) is given by
\begin{eqnarray}
    P_{\rm acc}^{{\bf Z}_n \rightarrow {\bf Z}_n'} = 
    \min \Bigg[ 1,\frac
    {Q({\bf Z}_n') \; P_{\rm gen}^{{\bf Z}_n' \rightarrow {\bf Z}_n} \;
	e^{-\frac{1}{2} \beta W'}}
    {Q({\bf Z}_n) \; P_{\rm gen}^{{\bf Z}_n \rightarrow {\bf Z}_n'} \;
	e^{-\frac{1}{2} \beta W}}
    \Bigg],
    \label{eq-Pacc}
\end{eqnarray}
where $P_{\rm gen}^{X \rightarrow Y}$ is the conditional
probability of generating a trial path $Y$ from existing path $X$.

For this study, we generate trial paths
by randomly choosing a ``shoot'' point
$\lambda_s$ along an existing path
(compare to Refs.\ \onlinecite{tps-review,dellago-rate,dellago-bd}).
Then, Langevin dynamics is used to propagate the system
from $\lambda_s \rightarrow 0$ (backward segment), followed by
$\lambda_s \rightarrow 1$ (forward segment).
Before running the backward segment, the velocities at the shoot point
must be reversed and then ordinary Langevin dynamics are used to
propagate the system \cite{tps-review}.
Once the trial path is complete,
all the velocities for the backward segment are reversed.
Since the stochastic Langevin algorithm is employed in the simulation,
it is not
necessary to perturb the configurational coordinates at the shoot point
to obtain a trial path that differs from the existing path.

The above recipe for generating trial paths leads to the following
statistical weights for the existing $Q({\bf Z}_n)$
and trial $Q({\bf Z}_n')$ paths
\begin{eqnarray}
    Q({\bf Z}_n) = e^{-\beta U_0(\vec{x}_0)} \;
	\prod_{i=0}^{n-1} 
	    p(\vec{x}_i \rightarrow \vec{x}_{i+1}), \nonumber \\
    Q({\bf Z}_n') = e^{-\beta U_0(\vec{x}_0')} \;
	\prod_{i=0}^{n-1} 
	    p(\vec{x}_i' \rightarrow \vec{x}_{i+1}'),
    \label{eq-Q}
\end{eqnarray}
where $p(\vec{x}_i \rightarrow \vec{x}_{i+1})$ is the
the transition probability for taking a dynamics step from configuration
$\vec{x}_i$ to $\vec{x}_{i+1}$ \cite{dellago-bd}.
We have assumed for simplicity that only one dynamics step is taken
at each value of $\lambda$; however, the approach allows for multiple steps.
The corresponding generating probabilities for the existing and trial
paths are given by
\begin{eqnarray}
    P_{\rm gen}^{{\bf Z}_n \rightarrow {\bf Z}_n'} = \nonumber \\
	p_{\rm choose} \; p_{\rm perturb} \;
	\prod_{i=s}^{n-1} p(\vec{x}_i' \rightarrow \vec{x}_{i+1}') \;
	\prod_{i=0}^{s-1} \bar{p}(\vec{x}_{i+1}' \rightarrow \vec{x}_i'),
    \nonumber \\
    P_{\rm gen}^{{\bf Z}_n' \rightarrow {\bf Z}_n} = \nonumber \\
	p_{\rm choose}' \; p_{\rm perturb}' \;
	\prod_{i=s}^{n-1} p(\vec{x}_i \rightarrow \vec{x}_{i+1}) \;
	\prod_{i=0}^{s-1} \bar{p}(\vec{x}_{i+1} \rightarrow \vec{x}_i),
    \label{eq-Pgen}
\end{eqnarray}
where $\bar{p}(\vec{x}_{i+1} \rightarrow \vec{x}_i)$ is the
transition probability of taking a {\it backward} step from $\vec{x}_{i+1}$
to $\vec{x}_i$. The ``bar'' notation is a reminder that the
velocities are reversed
for these segments. The probability of choosing a particular shoot point
$\lambda_s$ is denoted by $p_{\rm choose}$, and the probability of
a particular perturbation to the configurational coordinates
at the shoot point is given by $p_{\rm perturb}$.

Since we have chosen not to perturb the configurational
coordinates at the shoot point, and
any value of $\lambda$ along the path is equally likely to be chosen as the
shoot point, then $p_{\rm perturb}=p_{\rm perturb}'$ and
$p_{\rm choose}=p_{\rm choose}'$.
In addition, since the transition probabilities obey
detailed balance and preserve the canonical
distribution then \cite{dellago-lj}
\begin{eqnarray}
    \bar{p}(\vec{x}_{i+1} \rightarrow \vec{x}_i) =
	p(\vec{x}_i \rightarrow \vec{x}_{i+1}) \; 
	e^{-\beta \big(
	    U_{\lambda_{i+1}}(\vec{x}_i)-U_{\lambda_{i+1}}(\vec{x}_{i+1})
	\big)}.
    \label{eq-Pss}
\end{eqnarray}
Inserting Eqs.\ (\ref{eq-Q}), (\ref{eq-Pgen}) and (\ref{eq-Pss}) into
Eq.\ (\ref{eq-Pacc}) gives the acceptance criterion for trial paths
(compare to Eq.\ (45) in Ref.\ \onlinecite{athenes})
\begin{eqnarray}
    P_{\rm acc}^{{\bf Z}_n \rightarrow {\bf Z}_n'} = 
    \min \Bigg[ 1,
	{e^{-\beta \big( \delta W -\delta W' + \frac{1}{2}(W'-W) \big)}} 
    \Bigg],
    \label{eq-Pacc2}
\end{eqnarray}
where $\delta W$ is defined as the work accumulated up to the shoot point
for the existing path
\begin{eqnarray}
  \delta W=\sum_{i=0}^{s-1}\Bigl[
    U_{\lambda_{i+1}}(\vec{x}_i)-U_{\lambda_i}(\vec{x}_i)
    \Bigr].
  \label{eq-sepsw}
\end{eqnarray}
and $\delta W'$ is the equivalent quantity for the trial path.
Note that Eq.\ \eqref{eq-Pacc2} is independent of the details
of the fixed-$\lambda$ dynamics.

To clarify ambiguities in our original presentation of the SEPS approach
\cite{ytreberg-seps}, we also give details for applying it using overdamped
Langevin dynamics (i.e., Brownian dynamics).
In Ref.\ \onlinecite{ytreberg-seps}, backward segments were generated
using ordinary dynamics with {\it negative forces},
i.e., to be very clear, the force was taken to be identical to the
physical force, but opposite in sign.
Thus, the transition probabilities for forward and backward steps are 
approximately equal
\begin{eqnarray}
    \bar{p}(\vec{x}_{i+1} \rightarrow \vec{x}_i) \approx
	p(\vec{x}_i \rightarrow \vec{x}_{i+1}). \nonumber \\
	\text{(Brownian dynamics)} 
\end{eqnarray}
Equality occurs when the forces at $\vec{x}_i$ and $\vec{x}_{i+1}$
are identical.
The acceptance criterion becomes
\begin{eqnarray}
    P_{\rm acc}^{{\bf Z}_n \rightarrow {\bf Z}_n'} = 
    \min \Bigg[ 1,
	{e^{-\beta \big(
	    \frac{1}{2}(W'-W) + U_0(\vec{x}_0') - U_0(\vec{x}_0)
	\big)}}
    \Bigg]. \nonumber \\
    \text{(Brownian dynamics)}
\end{eqnarray}
Therefore, the criticism raised in a recent paper \cite{athenes}
is incorrect.

\subsection{\label{sec-bseps}Bennett averaging of path sampled work values}
The use of bi-directional data is worth considering for the SEPS method,\
just as it was for ordinary non-equilibrium Jarzynski work values.
Generalizing Bennett's method to include the
work values sampled from $D$ gives
\begin{eqnarray}
    {\sum_{N_{\rm f}}}^D
	\frac{e^{+\frac{1}{2}\beta W_{\rm f}}}
	{1 + e^{\beta \big( \eta + W_{\rm f} - \Delta F \big)}}
        \bigg[ {\sum_{N_{\rm f}}}^D e^{+\frac{1}{2} \beta W_{\rm f}} 
	    \bigg]^{-1} = \nonumber \\
    {\sum_{N_{\rm r}}}^D
	\frac{e^{+\frac{1}{2}\beta W_{\rm r}}}
	{1 + e^{\beta \big( -\eta + W_{\rm r} + \Delta F \big)}}
	\bigg[ {\sum_{N_{\rm r}}}^D e^{+\frac{1}{2} \beta W_{\rm r}} 
	    \bigg]^{-1}.
    \label{eq-bseps}
\end{eqnarray}
Thus, to obtain a Bennett-averaged estimate for \df, the path sampling
algorithm is applied to generate an ensemble of paths going from $U_0$ to
$U_1$ ($W_{\rm f}$, forward) and also for $U_1$ to $U_0$ ($W_{\rm r}$,
reverse). Then, Eq.\ \eqref{eq-bseps} is applied to the data.

\section{\label{sec-details}Simulation details}
To test the efficiency and precision of each method detailed above 
we use two relative solvation free energy calculations. One involves a 
large change in the van der Waals radius of a neutral particle in explicit
solvent (``growing''), and the other is a large change 
in the charge of the particle while keeping the size fixed (``charging'').

The system used in both cases consists of a single Lennard-Jones particle in
a 24.93 \AA\ box of 500 TIP3P water molecules. For all simulations,
the molecular simulation package TINKER 4.2 was used \cite{tinker}.
The temperature of the system was maintained at 300.0 K using 
Langevin dynamics with a friction coefficient of 5.0 $\rm ps^{-1}$. 
RATTLE was used to constrain all hydrogens to their ideal lengths
\cite{rattle}, allowing a 2.0 fs time step. A cutoff of 12.465 \AA\ was
chosen for electrostatic and van-der-Waals interactions with a smoothing
function implemented from 10.465 to 12.465 \AA. It is expected
that the use of cutoffs will introduce systematic errors into
the \df\ calculation, however, in this report we are only interested
in comparing \df\ methodologies---we do not compare our results
to experimental data.

For the first test case, a neutral Lennard-Jones particle was
``grown'' from 2.126452 \AA\ to 6.715999 \AA.
The sizes were chosen to be that of lithium and cesium from the
OPLS-AA forcefield \cite{oplsaa}.
In the second test case, the Lennard-Jones particle remains at a fixed size of
2.126452 \AA, but the charge is changed from -e/2 to +e/2.
For each test case, and each \df\ method, the system was initially
equilibrated for 100 ps ($ 5 \times 10^4$ dynamics steps).
The initial equilibration is not included in the 
total computational time listed in the results,
however, since every method was given 
identical initial equilibration times, the efficiency analysis is fair.

The $\lambda$-scaling (i.e., the form of the hybrid potential $U_{\lambda}$)
used for all \df\ methods in this study
was chosen to be the default implementation
within the TINKER package \cite{tinker}.
If a particle's charge is varied from $q_0$ to $q_1$, the 
hybrid potential is simply the regular potential energy calculated using a
hybrid charge of
\begin{eqnarray}
    q_{\lambda} = \lambda q_1 + (1-\lambda) q_0.
    \label{eq-Ulq}
\end{eqnarray}
Similarly, if a particle has a change in the van der Waals parameters
$r, \; \epsilon$ the hybrid parameters are given by
\begin{eqnarray}
    r_{\lambda} = \lambda r_1 + (1-\lambda) r_0,\nonumber \\
    \epsilon_{\lambda} = \lambda \epsilon_1 + (1-\lambda) \epsilon_0.
    \label{eq-Ulr}
\end{eqnarray}
The free energy slope as a function of $\lambda$
for both the growing and charging test cases are shown in Figs.\
\ref{fig-grow_profile} and \ref{fig-charge_profile}. The smoothness of
both plots suggests that a more sophisticated $\lambda$-scaling is not
necessary for this study. If, for example, we had chosen to grow a
particle from nothing, then it is likely that a different scaling
would be needed 
(such as in Refs.\ 
\onlinecite{brooks-ldyn,karplus-jcp,shirts-jcp,shirts-benn}).

\subsection{Thermodynamic integration calculations}
For thermodynamic integration (TI), equilibrium simulations were performed
at each value of $\lambda$. An equal amount of simulation time
was devoted to each of 21 equally spaced values of
$\lambda=0.0, 0.05, 0.1,...,0.9, 0.95, 1.0$.
Averages of the slope
${\rm d}F/{\rm d}\lambda = \langle {\rm d}U/{\rm d}\lambda \rangle_\lambda$,
shown in Figs.\ \ref{fig-grow_profile} and \ref{fig-charge_profile},
were collected for each value of $\lambda$.
The first 50\% of the slope data were discarded for
equilibration. Finally, the data were used to estimate the integral
in Eq.\ (\ref{eq-ti}) using the trapezoidal rule.
Note that higher order integration schemes were also attempted,
but did not change the results, suggesting that the curves in
Figs.\ \ref{fig-grow_profile} and \ref{fig-charge_profile}
are smooth enough that high order integration schemes are not
needed for this report. Also, the percentage of data that was
discarded for equilibration was varied from 25-75\% with no
significant changes to the results.

\subsection{Adaptive integration calculations}
Adaptive integration (AIM) results were obtained by collecting
the slope of the free energy
${\rm d}F/{\rm d}\lambda = \langle {\rm d}U/{\rm d}\lambda \rangle_\lambda$,
by starting the simulation from an equilibrated
configuration at $\lambda=0$ and
performing one dynamics step. Immediately following the single
step, a Monte Carlo move in $\lambda$ was attempted, which
was accepted with probability given by Eq.\ (\ref{eq-Paim}).
The pattern of one dynamics step followed by one Monte Carlo trial move
was repeated until a total of $10^6$ dynamics steps (and thus
$10^6$ Monte Carlo attempts) had been performed.
The same $\lambda$ values used in TI are also used for AIM, thus
$\lambda=0.0, 0.05, 0.1,...,0.9, 0.95, 1.0$ are the only allowed
values.
For this report Monte Carlo moves were
attempted between neighboring values of $\lambda$ only, i.e., a move
from $\lambda$=0.35 to 0.4 or 0.3 may be attempted but not to 0.45.
Also, all $\delta \hat{F}(\lambda_i)$ values of Eq.\ (\ref{eq-dFaim}) were
initially set to zero.
The estimate of the free energy was obtained by numerically approximating
the integral in Eq.\ (\ref{eq-ti}) using the trapezoidal rule. As with
TI, higher order integration schemes did not change the results.

\subsection{Free energy perturbation and equilibrium Bennett calculations}
All free energy perturbation calculations (forward Eq.\ \eqref{eq-fepf}
and reverse Eq.\ \eqref{eq-fepr}),
and equilibrium Bennett computations (Eq.\ \eqref{eq-benn})
were performed on the same set of configurations as for TI.
Specifically, equilibrium simulations were performed
at each of 21 equally spaced values of
$\lambda=0.0, 0.05, 0.1,...,0.9, 0.95, 1.0$,
and the first 50\% of the data were discarded for
equilibration.

\subsection{Jarzynski estimate calculations}
Estimates of the free energy using the non-equilibrium work values
were computed using Eq.\ (\ref{eq-jarz}) for Jarzynski averaging,
and Eq.\ (\ref{eq-bjarz}) for Bennett averaging. ``Forward''
non-equilibrium paths were generated by starting the
simulation from an equilibrated
configuration at $\lambda=0$, then incrementing the value of
$\lambda$, followed by another dynamics step, and so on until $\lambda=1$.
Thus, only
one dynamics step was performed at each value of $\lambda$. The
work value associated with the path was then computed using
Eq.\ (\ref{eq-Wjarz}). Between each path, the system
was simulated for 100 dynamics steps at $\lambda=0$, starting
with the last $\lambda=0$ configuration---thus the $\lambda=0$
equilibrium ensemble was generated ``on the fly.''

Similarly, ``reverse'' non-equilibrium paths were generated by starting
each simulation from configurations in the $U_1$ equilibrium ensemble
and switching from $\lambda=1$ to $\lambda=0$.

\subsection{\label{sec-sepsdetails}Single-ensemble path sampling calculations}
For the single-ensemble path sampling (SEPS)
method, we first generated an initial path using standard Jarzynski
formalism. The only difference between the paths described above and
the initial path for SEPS
was that, due to the computer memory needed to store a path,
the number of $\lambda$-steps was limited to 500 for this study.
In other words,
if the desired path should contain around 2000 dynamics steps,
the simulation would perform four dynamics steps at each $\lambda$
value giving a total simulation time of 1996 dynamics steps for each path
(note that simulation at $\lambda=1$ was not necessary).

\begin{table*}[t]
\begin{tabular}{lccccccccc}
    \hline \hline
    Steps & AIM & TI & SEPS & BSEPS & Jarz &
	BJarz & Benn & FEPF & FEPR\\
    \hline
    2E3 & 16.3(4.6) & 16.5(6.1) & --- & --- & ---
	& --- & 16.7(6.2)& 18.7(6.7)& 14.5(5.7) \\
    \hline
    4E3 & 14.4(3.9) & 13.2(4.4) & --- & --- & ---
	& --- & 13.4(4.4) & 14.7(4.7) & 11.9(4.2) \\
    \hline
    9E3 & 10.4(3.3) & 11.2(3.6) & --- & --- & 7.9(1.3)
	& --- & 11.3(3.6) & 12.3(3.9) & 10.1(3.3) \\
    \hline
    1.7E4 & 8.94(2.35) & 9.7(2.46) & --- & --- & 7.56(0.93)
	& 7.53(1.13) & 9.75(2.46) & 10.48(2.70) & 8.92(2.26) \\
    \hline
    3.5E4 & 7.51(0.52) & 8.32(1.35) & --- & --- & 7.62(0.84)
	& 7.47(0.71) & 8.36(1.38) & 8.91(1.63) & 7.74(1.11) \\
    \hline
    7E4 & 7.38(0.48) & 7.89(1.17) & --- & --- & 7.55(0.67)
	& 7.38(0.59) & 7.92(1.19) & 8.35(1.40) & 7.46(0.97) \\
    \hline
    1.3E5 & 7.35(0.36) & 7.18(0.65) & 7.15(0.79) & --- & 7.34(0.49) &
	7.36(0.38) & 7.22(0.64) & 7.56(0.68) & 6.83(0.68) \\
    \hline
    2.7E5 & 7.34(0.23) & 7.19(0.22) & 7.19(0.62) & 6.95(0.56) & 7.35(0.44) 
	& 7.28(0.24) & 7.21(0.22) & 7.29(0.25) & 7.08(0.20) \\
    \hline
    5.5E5 & 7.22(0.12) & 7.18(0.11) & 7.19(0.29) & 7.12(0.46) & 7.32(0.28)
	& 7.23(0.20) & 7.18(0.12) & 7.22(0.11) & 7.16(0.13) \\
    \hline
    1E6 & 7.19(0.07) & 7.26(0.18) & 7.17(0.18) & 7.23(0.20) & 7.25(0.23)
	& 7.22(0.14) & 7.26(0.18) & 7.28(0.18) & 7.24(0.20) \\
    \hline \hline
    \end{tabular}
    \caption{
	Free energy difference estimates obtained for
	changing the Lennard-Jones size
	of a neutral particle in a box of explicit water. Results are shown
	for various methods described in the text as a function of
	the number of dynamics steps used in the simulation.
	Table entries are the mean estimates from 16 independent simulations
	with the standard deviation shown in parentheses.
	For single-ensemble path sampling (SEPS and BSEPS)
	and Jarzynski methods (Jarz and BJarz),
	only the most efficient results are shown.
	The table shows that in the limit of long simulation times
	($10^6$ dynamics steps) all methods produce average
	\df\ estimates that roughly agree. The table
	also shows that AIM provides the most precise long-simulation
	estimate.
    \label{tab-grow1}
    }
\end{table*}

Once an initial path was generated as described above,
a trial path was created by perturbing the
old path as described in Sec.\ \ref{sec-seps}. Then, the new path was
accepted with probability given by Eq.\ (\ref{eq-Pacc2}).
Importantly, if the new path was rejected, then the old path
was counted again in the path ensemble.
Also, as with any Monte Carlo approach,
an initial equilibration phase was needed.
For this report, the necessary amount of equilibration
was determined by studying the dependence of the average free energy
estimate, after $10^6$ dynamics steps, from 16
independent trials, as a function of the number
of paths that were discarded for equilibration. The optimal number
of discarded paths was then chosen to be where the average free
energy estimate no longer depends on the number of discarded paths.

\section{\label{sec-results} Results and Discussion}
Using the simulation details described above, two relative
solvation free energy calculations were carried out in a box of
500 TIP3P water molecules.
Each of the free energy methods described above were used to
estimate \df. Specifically, we compare:
\begin{itemize}
    \item adaptive integration (AIM) using Eqs.\ \eqref{eq-ti} and
	\eqref{eq-Paim};
    \item thermodynamic integration (TI) using Eq.\ \eqref{eq-ti};
    \item uni-directional single-ensemble path sampling (SEPS) using
	Eq.\ \eqref{eq-seps};
    \item bi-directional single-ensemble path sampling with Bennett averaging
	(BSEPS) using Eq.\ \eqref{eq-bseps};
    \item uni-directional Jarzynski averaging of work values (Jarz)
	using Eq.\ \eqref{eq-jarz};
    \item bi-directional Bennett averaging of Jarzynski work values
	(BJarz) using Eq.\ \eqref{eq-bjarz};
    \item Equilibrium Bennett approach (Benn) using Eq.\ \eqref{eq-benn}; and
    \item multi-stage free energy perturbation in the forward (FEPF) and 
	reverse (FEPR) directions, using, respectively Eqs.\ \eqref{eq-fepf}
	and \eqref{eq-fepr}.
\end{itemize}

\subsection{\label{sec-growresults}Growing a Lennard-Jones particle}
We first compute the free energy required to grow a neutral
particle from 2.126452 \AA\ to 6.715999 \AA\ in 500 TIP3P waters.

\begin{figure}
    \begin{center}
	\includegraphics[scale=0.3,clip]{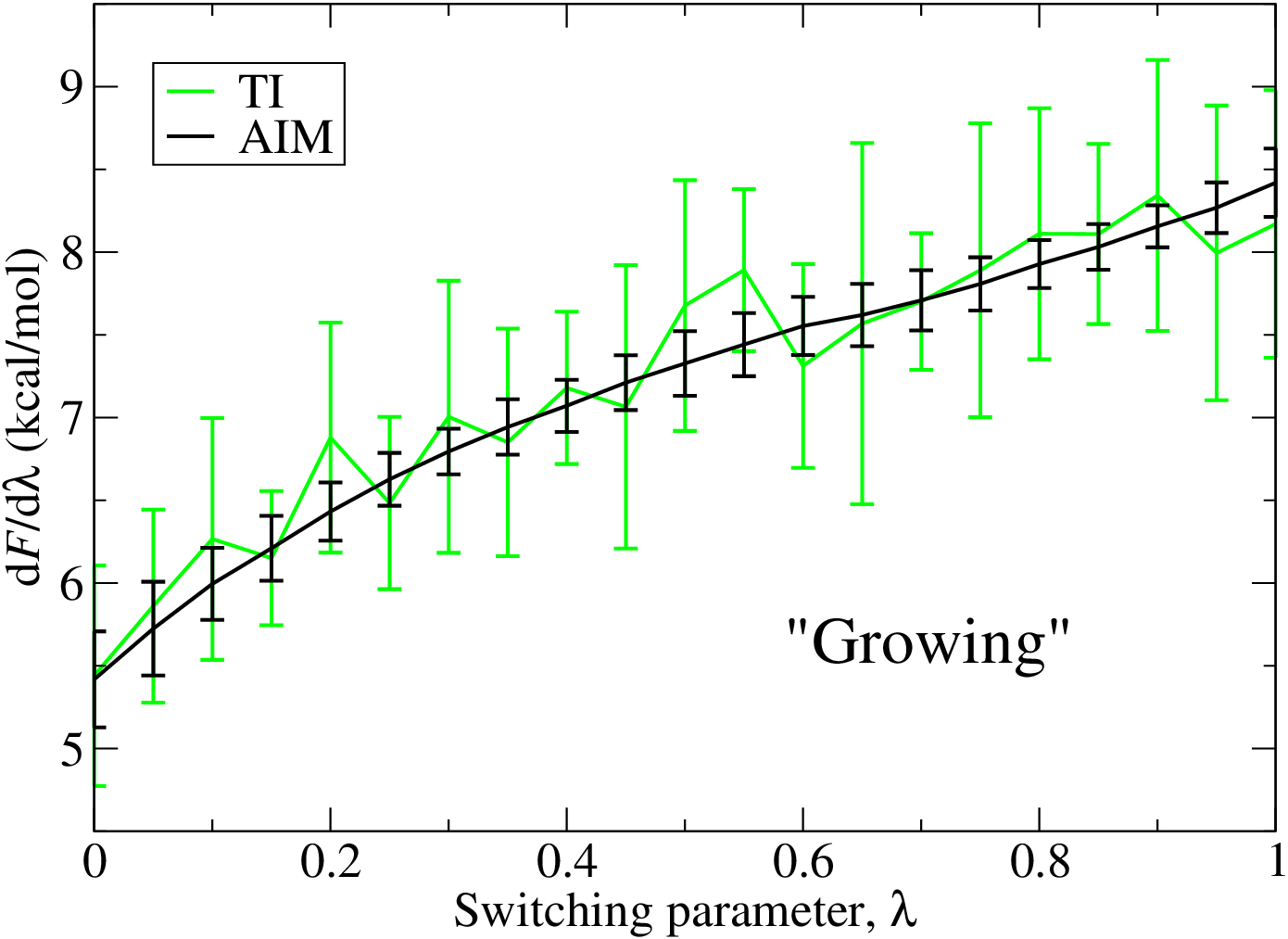}
    \end{center}
    \caption{\label{fig-grow_profile}
	The slope of the free energy d$F$/d$\lambda$ as a function
	of $\lambda$ for changing the Lennard-Jones size
	of a neutral particle in a box of explicit water. Results for both 
	TI and AIM methods are shown for $10^6$ dynamics steps.
	The data show the averages (data points) and standard deviations
	(error bars) from 16 independent simulations for each method.
	The figure demonstrates that AIM has the ability to
	sample the $\lambda$-path more efficiently,
	thus producing a much smoother and more precise
	profile compared to TI.
	Thus, AIM is preferred over TI for computing the potential
	of mean force for this system.
	In addition, the smoothness of the profile suggests that the
	switching function $U_{\lambda}$ of Eq.\ (\ref{eq-Ulr})
	used in this report is adequate.
    }
\end{figure}

Figure \ref{fig-grow_profile} shows the slope of the free energy
(${\rm d}F/{\rm d}\lambda = \langle {\rm d}U/{\rm d}\lambda \rangle_\lambda$)
as a function of $\lambda$
for both TI and AIM after $10^6$ Langevin dynamics steps.
The figure suggests that AIM can more efficiently sample the profile.
In AIM, configurations
are not forced to remain at a particular $\lambda$, but may switch to
another value of $\lambda$ if it is favorable to do so.
Such ``cross-talk'' is apparently the source of the
smoother $\lambda$-profile compared to TI.
Table \ref{tab-grow1} shows \df\ estimates for the different approaches
used in this report. Note that for all non-equilibrium
approaches, only the most efficient data are shown.
For SEPS and BSEPS all paths were composed of 500 $\lambda$-steps
(restricted to 500 due to computer memory)
with 40 dynamics steps at each value of $\lambda$.
For Jarz and BJarz the paths were composed of
10 000 $\lambda$-steps with one dynamics step
at each value of $\lambda$. For all of these non-equilibrium
data, the standard deviation of the work values were
$\sigma_W \approx 0.8 \; {\rm kcal/mol} \approx 1.3 \; k_BT$,
in agreement with previous studies \cite{hummer,crooks-pre}.
At least five different path lengths were attempted for each
non-equilibrium method to determine the most efficient.

Table \ref{tab-grow1} demonstrates that, for long simulation times, all
methods produce roughly the same average \df\ estimate.
Also, the table clearly shows that,
given $10^6$ dynamics steps, AIM provides the most
precise free energy estimates.

\begin{table}
\begin{tabular}{lcc}
    \hline \hline
    Method & \; Within 1.0 kcal/mol & \; Within 0.5 kcal/mol \\
    \hline
    AIM   & 23 000   & 30 000 \\
    \hline
    TI    & 89 000   & 181 000 \\
    \hline
    SEPS  & 140 000  & 377 000 \\
    \hline
    BSEPS & 279 000  & 444 000 \\
    \hline
    Jarz  & 18 000   & 127 000 \\
    \hline
    BJarz & 26 000   & 96 000 \\
    \hline
    Benn  & 90 000   & 180 000 \\
    \hline
    FEPF  & 104 000  & 191 000 \\
    \hline
    FEPR  & 60 000   & 184 000 \\
    \hline \hline
    \end{tabular}
    \caption{
	Number of dynamics steps necessary to be within a specified tolerance
	of the correct result $\Delta F_{\rm long \; sim}$ = 7.23 kcal/mol,
	average \df\ estimate
	at $10^6$ dynamics steps for all methods, for growing a
	Lennard-Jones particle in explicit solvent.
	The first column is the method used to obtain the estimate.
	The second column is the number of dynamics steps needed to estimate
	\df\ within 1.0 kcal/mol of $\Delta F_{\rm long \; sim}$ with
	an uncertainty less than 1.0 kcal/mol.
	The third column is the number of
	dynamics steps needed to obtain an estimate within 0.5 kcal/mol
	with an uncertainty less than 0.5 kcal/mol.
    \label{tab-grow2}
    }
\end{table}

Table \ref{tab-grow2} shows the approximate number of
dynamics steps needed by each
method to obtain a free energy estimate within a specific tolerance
of $\Delta F_{\rm long \; sim}$ (average of all estimates at $10^6$
dynamics steps). Note that the number of dynamics steps needed
for the SEPS and BSEPS methods are large due to
the fact that whole paths must be discarded for
equilibration of the path ensemble.
For all methods except AIM, the table entries for
Table \ref{tab-grow2} were estimated using
linear interpolation of the data in Table \ref{tab-grow1}.
From the data in Table \ref{tab-grow2}, if
the desired precision is less than 1.0 kcal/mol, then
AIM, Jarz and BJarz appear to be the best methods. However, if
the desired precision is less than 0.5 kcal/mol, then AIM is
the best choice.

Tables \ref{tab-grow1} and \ref{tab-grow2}, taken together, demonstrate the
difference between using equilibrium data in the ``forward'' (FEPF) and
``reverse'' (FEPR) directions.
While, the results are similar for $10^6$ dynamics steps, it is clear that
FEPR produces the desired results more rapidly than FEPF indicating
that the configurational overlap is greater in the reverse direction.
However, the FEPR data also tends to ``overshoot'' the correct value by
a small margin which makes convergence of the FEPR estimate difficult
to judge. 

Thus, we conclude that, for growing a Lennard-Jones
particle in explicit solvent, the preferred method depends upon the
type of estimate one wishes to generate. If a very precise high-quality
estimate is desired, then AIM is the best choice by a considerable margin.
If a very rapid estimate of \df, with an uncertainty of less
than 1.0 kcal/mol, is desired, then then comparable results are seen using
AIM, Jarz and BJarz methodologies. If the \df\ estimate is to be within
0.5 kcal/mol, then AIM is the best choice.

Finally, if the desired result is the potential of mean force, then AIM
will generate a much smoother curve than TI.

\begin{figure*}
    \begin{center}
	\includegraphics[scale=0.3,clip]{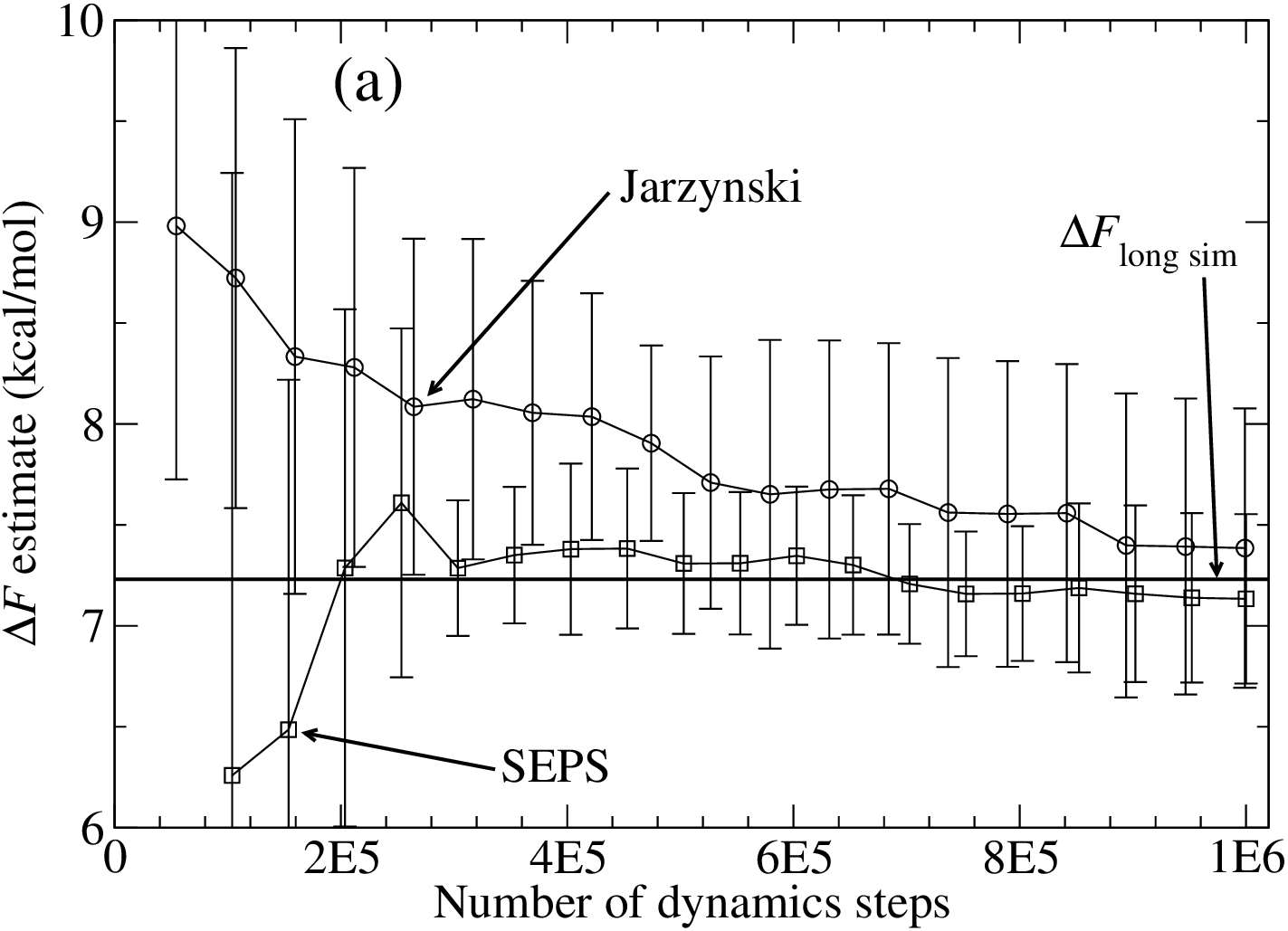}
	\includegraphics[scale=0.3,clip]{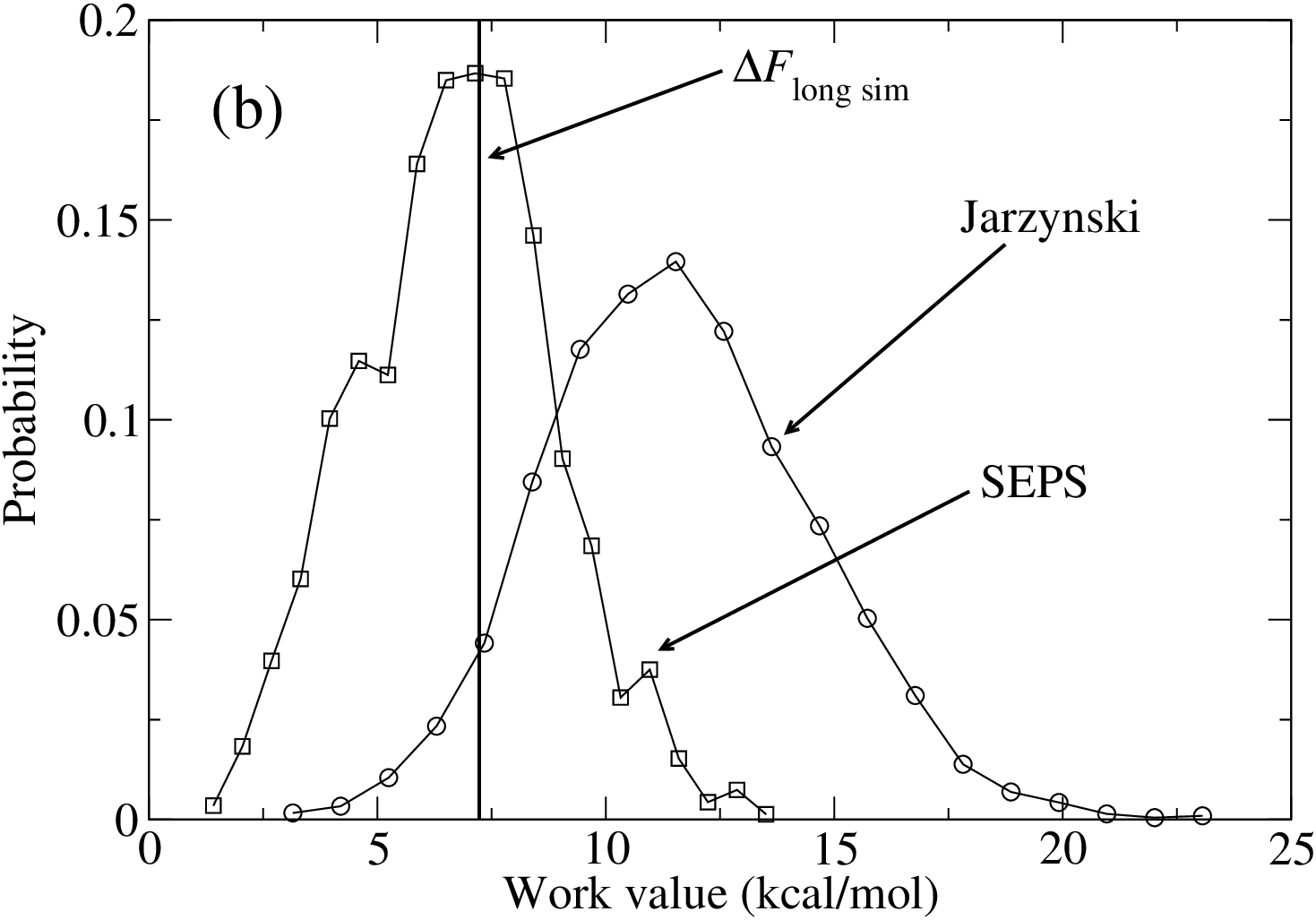}
    \end{center}
    \caption{\label{fig-fast}
	(a) ``Fast-growth'' uni-directional free energy difference estimates
	obtained for changing the Lennard-Jones size
	of a neutral particle in a box of explicit water. Results are shown
	for both SEPS and Jarz methods as a function of
	the number of dynamics steps used in the simulation. For both methods,
	fast-growth work values were generated by simulating roughly
	2000 dynamics steps per path, which is ten times shorter
	than optimal.
	The solid horizontal line represents the best
	estimate of the free energy difference $\Delta F_{\rm long \; sim}$
	based on averaging all results shown in Table \ref{tab-grow1}
	at $10^6$ dynamics steps.
	The averages (data points) and standard deviations (errorbars) are
	from 16 independent simulations.
	(b) Histograms of the work values used to generate the free
	energy estimates for both the SEPS and Jarz methods.
	The plots demonstrate the potential usefulness of using
	path sampling over regular Jarzynski averaging. Specifically,
	if the work values are fast-growth and uni-directional, then
	SEPS is able to bias the work values in such a way to improve
	the free energy estimate.
	Note that for all the SEPS data shown, the first 50 work values
	are thrown away for equilibration as, described in Sec.\
	\ref{sec-sepsdetails}.
    }
\end{figure*}

\subsubsection{Fast-growth uni-directional data}
We now consider non-equilibrium uni-directional fast-growth
data, i.e., generated by switching the system rapidly from $U_0$ (small
particle) to $U_1$ (large particle).
Importantly, there will be an advantage to generating uni-directional
data in some cases, since only the $U_0$ equilibrium ensemble is
needed to estimate \df.

In contrast to the data shown in Tables \ref{tab-grow1} and \ref{tab-grow2},
where the lengths of the non-equilibrium switching trajectories were
pre-optimized, here we focus on the efficacy of the methods using
non-optimal, rather fast switching.
After all, when attempting a free energy computation
on a new system, there is no way to know in advance the optimal path length
(number of $\lambda$-steps).
Substantial optimization may be
needed for both SEPS and Jarz methods to work efficiently.

Here, we test the SEPS and Jarz methods using short paths with
an equal number of dynamics steps.
For SEPS, 500 $\lambda$-steps with four dynamics steps at each
value of $\lambda$ was used, producing a distribution of work values
with $\sigma_W=2.1$ kcal/mol.
For Jarz, 2000 $\lambda$-steps with one
dynamics step at each value of $\lambda$ was used, producing a distribution
of work values with $\sigma_W=2.9$ kcal/mol.
Note that these paths are roughly ten times
shorter than optimal and thus $\sigma_W$ is 3-4 times larger
than the optimal value of $\sim k_BT$.

Figure \ref{fig-fast} gives a comparison between SEPS and Jarz
methods for the fast-growth uni-directional paths just described.
The upper figure
(a) shows the average free energy estimates and standard
deviations for both the SEPS and Jarz methods. The lower
figure (b) gives the histogram of the work values for each method.
Both figures also show the ``correct'' value $\Delta F_{\rm long \; sim}$,
generated from a very long simulation.
The figures clearly demonstrate that, for fast-growth data,
SEPS has the ability to ``shift'' the work values such that the
\df\ value is near the center of the
work value distribution---rather than in the tail of the
distribution as with the Jarz method.
Thus, the SEPS results converge more rapidly than Jarz to the correct
value of \df.

We suggest that the the SEPS method may find the greatest use
for the ability to bias fast-growth work values to obtain the correct
value of \df, as shown here.

\begin{figure}
    \begin{center}
	\includegraphics[scale=0.3,clip]{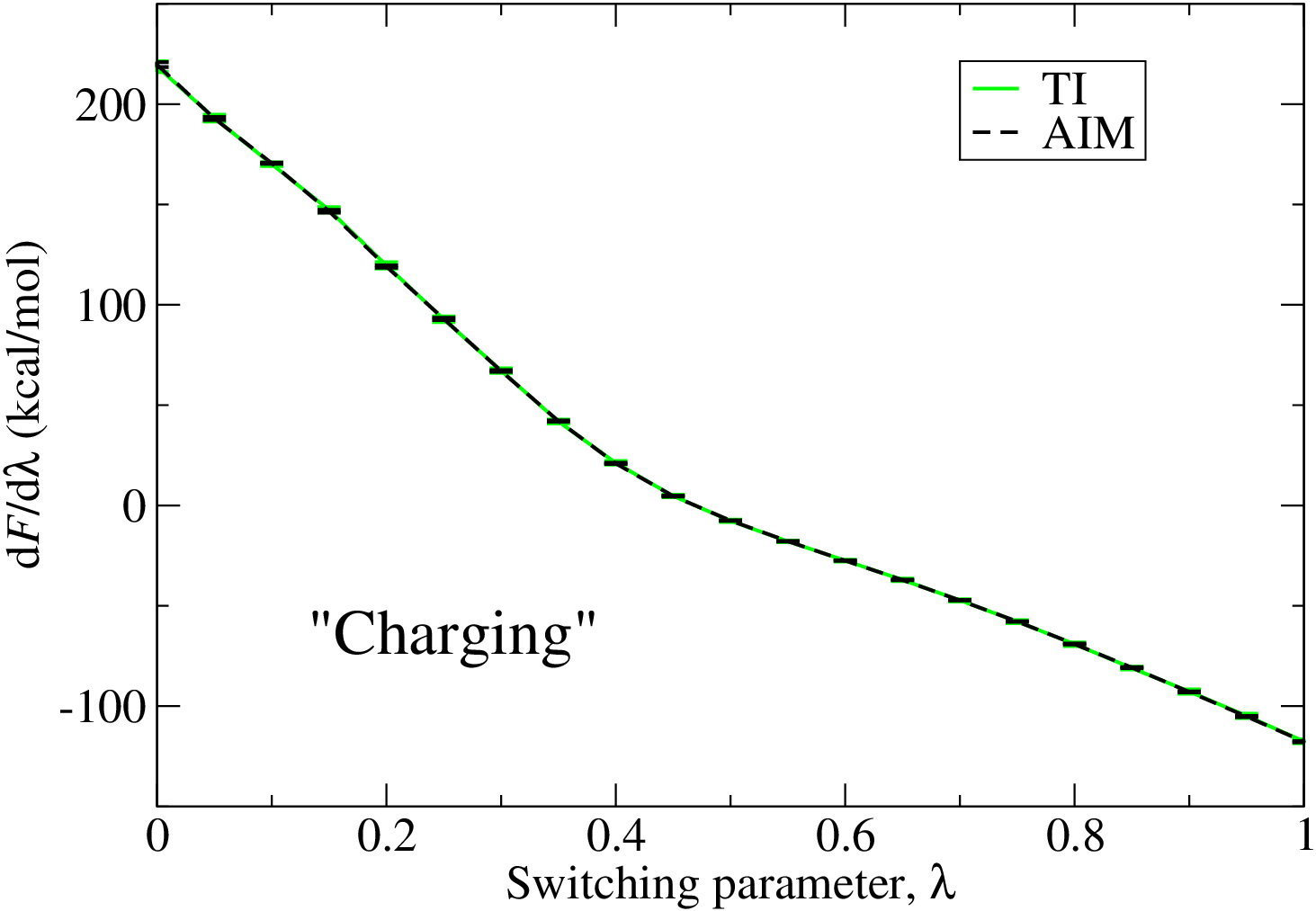}
    \end{center}
    \caption{\label{fig-charge_profile}
	The slope of the free energy d$F$/d$\lambda$ as a function
	of $\lambda$ for a changing the charge of a
	Lennard-Jones particle in a box of explicit water from
	-e/2 to +e/2. Results for both TI and AIM
	methods are shown for $10^6$ dynamics steps.
	The data show the averages (data points) and standard deviations
	(error bars) from 16 independent simulations for each method.
	The errorbars are too small to resolve on the plot shown, however,
	it should be noted that the average uncertainty in the the slope
	for AIM is 0.38 kcal/mol and for TI is 1.05 kcal/mol, suggesting
	that AIM has the ability to produce a more precise
	profile compared to TI.
	Thus, AIM is preferred over TI for computing the potential
	of mean force for this system.
	The smoothness of the profile also suggests that the
	switching function $U_{\lambda}$ of Eq.\ (\ref{eq-Ulq})
	used in this report is adequate.
    }
\end{figure}

\subsection{Charging a Lennard-Jones particle}
We next compute the free energy
required to charge a particle from -e/2 to +e/2
in 500 TIP3P waters.

\begin{table*}[t]
\begin{tabular}{lccccccccc}
    \hline \hline
    Steps & AIM & TI & SEPS & BSEPS & Jarz &
	BJarz & Benn & FEPF & FEPR\\
    \hline
    2E3 & 8.5(5.5) & 24.5(2.3) & --- & --- & ---
	& --- & 24.4(2.3) & 28.7(2.8) & 20.0(2.1) \\
    \hline
    4E3 & 9.7(6.6) & 21.5(3.0) & --- & --- & ---
	& --- & 21.4(3.1) & 25.4(3.0) & 17.7(3.1) \\
    \hline
    9E3 & 14.6(11.4) & 20.1(1.7) & --- & --- & ---
	& --- & 20.1(1.8) & 22.6(1.8) & 17.6(2.1) \\
    \hline
    1.7E4 & 18.6(10.8) & 18.5(1.2) & --- & --- & ---
	& --- & 18.5(1.2) & 20.3(1.1) & 16.8(1.4) \\
    \hline
    3.5E4 & 19.7(4.6) & 18.44(0.87) & --- & --- & 19.15(0.70)
	& 18.42(0.74) & 18.39(0.90) & 19.56(1.05) & 17.34(0.70) \\
    \hline
    7E4 & 18.42(0.43) & 18.38(0.69) & --- & --- & 18.82(0.61)
	& 18.29(0.40) & 18.33(0.69) & 19.18(0.87) & 17.64(0.69) \\
    \hline
    1.3E5 & 18.41(0.26) & 18.34(0.71) & --- & --- & 18.72(0.55) &
	18.20(0.46) & 18.28(0.72) & 18.76(0.83) & 17.78(0.80) \\
    \hline
    2.7E5 & 18.27(0.21) & 18.35(0.45) & 18.47(1.03) & 18.23(0.59) & 18.55(0.42) 
	& 18.16(0.29) & 18.29(0.45) & 18.62(0.54) & 18.09(0.46) \\
    \hline
    5.5E5 & 18.26(0.13) & 18.28(0.28) & 18.25(0.49) & 18.43(0.43) & 18.44(0.32)
	& 18.13(0.19) & 18.20(0.29) & 18.28(0.39) & 18.25(0.26) \\
    \hline
    1E6 & 18.23(0.13) & 18.28(0.30) & 18.23(0.30) & 18.30(0.42) & 18.32(0.26)
	& 18.18(0.16) & 18.21(0.31) & 18.20(0.33) & 18.25(0.31) \\
    \hline \hline
    \end{tabular}
    \caption{
	Free energy difference estimates obtained for
	changing the charge of a Lennard-Jones particle from
	-e/2 to +e/2 in a box of explicit water. Results are the
	averages from 16 independent simulations
	for various methods described in the text as a function of
	the number of dynamics steps used in the simulation.
	The standard deviation is shown in parentheses.
	For single-ensemble path sampling (SEPS and BSEPS)
	and Jarzynski methods (Jarz and BJarz),
	only the most efficient results are shown.
	The table shows that in the limit of long simulation times
	($10^6$ dynamics steps) all methods produce average
	\df\ estimates that roughly agree. The table
	also shows that AIM and BJarz approaches
	provide the most precise long-simulation
	estimate.
    \label{tab-charge1}
    }
\end{table*}

Figure \ref{fig-charge_profile} shows the slope of the free energy
(${\rm d}F/{\rm d}\lambda = \langle {\rm d}U/{\rm d}\lambda \rangle_\lambda$)
as a function of $\lambda$
for both TI (green) and AIM (black) after $10^6$ Langevin dynamics steps.
The data shown in the plot are the mean (data points) and standard
deviation (errorbars) for 16 independent trials.
While the errorbars are too small to resolve on the plot shown,
the average uncertainty in the the slope for AIM is 0.38 kcal/mol
and for TI is 1.05 kcal/mol, suggesting
that AIM has the ability to produce more precise
slope data compared to TI.

Table \ref{tab-charge1} shows \df\ estimates for the different approaches.
For all non-equilibrium approaches, only the most efficient data are shown.
For SEPS and BSEPS the paths were composed of
500 $\lambda$-steps
(restricted to 500 due to computer memory)
with 80 dynamics steps at each value of $\lambda$.
For Jarz the paths were composed of
40 000 $\lambda$-steps with one dynamics step
at each value of $\lambda$, and
for BJarz, 20 000 $\lambda$-steps with one
dynamics step at each value of $\lambda$ were used.
For all of these non-equilibrium data, the standard
deviation of the work values were
$\sigma_W \approx 0.8 \; {\rm kcal/mol} \approx 1.3 \; k_BT$,
in agreement with previous studies \cite{hummer,crooks-pre}, and with
the growing data in this study.
At least four different path lengths were attempted for each
non-equilibrium method to determine the most efficient.

Table \ref{tab-charge1} demonstrates that, for long simulation times, all
methods produce roughly the same average \df\ estimate.
Also, the table shows that, given $10^6$ dynamics steps,
AIM and BJarz methodologies provide the most
precise free energy estimates.

Tables \ref{tab-charge1} and \ref{tab-charge2} show the
difference between using equilibrium data in the ``forward'' (FEPF) and
``reverse'' (FEPR) directions.
While, the results are similar for $10^6$ dynamics steps, it is clear that
FEPF produces the desired results more rapidly than FEPR indicating
that the configurational overlap is greater in the forward direction.
However, the FEPF data also tends to ``overshoot'' the correct value by
a small margin which makes convergence of the FEPF estimate difficult
to judge.

\begin{table}
\begin{tabular}{lcc}
    \hline \hline
    Method & \; Within 1.0 kcal/mol & \; Within 0.5 kcal/mol \\
    \hline
    AIM   & 52 000   & 64 000 \\
    \hline
    TI    & 27 500   & 243 000 \\
    \hline
    SEPS  & 291 000  & 515 000 \\
    \hline
    BSEPS & 399 000  & 487 000 \\
    \hline
    Jarz  & 40 000   & 180 000 \\
    \hline
    BJarz & 40 000   & 69 000 \\
    \hline
    Benn  & 29 000   & 245 000 \\
    \hline
    FEPF  & 43 000   & 335 000 \\
    \hline
    FEPR  & 26 000   & 252 000 \\
    \hline \hline
    \end{tabular}
    \caption{
	Number of dynamics steps necessary to be within a specified tolerance
	of the correct result $\Delta F_{\rm long \; sim}$ = 18.24 kcal/mol,
	average \df\ estimate
	at $10^6$ dynamics steps for all methods, for charging a
	Lennard-Jones particle in explicit solvent.
	The first column is the method used to obtain the estimate.
	The second column is the number of dynamics steps needed to estimate
	\df\ within 1.0 kcal/mol of $\Delta F_{\rm long \; sim}$ with
	an uncertainty less than 1.0 kcal/mol.
	The third column is the number of
	dynamics steps needed to obtain an estimate within 0.5 kcal/mol
	with an uncertainty less than 0.5 kcal/mol.
    \label{tab-charge2}
    }
\end{table}

For fast estimation of free energy differences,
Table \ref{tab-charge2} shows the number
of dynamics steps needed by each method to obtain a free energy
estimate within a specific tolerance
of $\Delta F_{\rm long \; sim}$ (average of all estimates at $10^6$
dynamics steps). Note that the number of dynamics steps needed for
the SEPS and BSEPS methods are large due to
the fact that many paths must be discarded for
equilibration of the path ensemble.
For all methods except AIM, the entries in Table \ref{tab-charge2}
were estimated using
linear interpolation of the data in Table \ref{tab-charge1}.
From the data in the table, if the desired precision is less
than 1.0 kcal/mol, then all methods other than SEPS and BSEPS
produce comparable results. However, if
the desired precision is less than 0.5 kcal/mol, then AIM and BJarz
approaches are best.

We conclude that, when charging a Lennard-Jones
particle in explicit solvent, the preferred methodology
depends upon the type of estimate one wishes to generate.
If a very high quality estimate is desired, then AIM is the best
choice, closely followed by BJarz.
If a very rapid estimate of \df, with an uncertainty of less
than 1.0 kcal/mol, is desired, then then comparable results are seen using
all methodologies {\it except} for SEPS and BSEPS.
If the \df\ estimate is to be within
0.5 kcal/mol, then AIM and BJarz are the best choices.

Finally, if the desired result is the potential of mean force, then AIM
will generate a much smoother curve than TI.

\subsection{A second look at a two-dimensional model}
Because SEPS proved orders of magnitude more efficient than TI
and Jarz in the study of a two-dimensional model
\cite{ytreberg-seps}, we return to that model in an effort to
understand the decreased effectiveness of SEPS
in the present study. Specifically, we use the model from
Ref.\ \onlinecite{ytreberg-seps}, but now for a wide range of
conformational sampling barrier
heights (fixed $\lambda$), and
then compare SEPS to TI, as in our original study.
Note, that we use the term
``conformational sampling barrier'' to distinguish it from the barrier in
$\lambda$-space.

Some alterations to our approach in Ref.\ \onlinecite{ytreberg-seps}
were necessary to provide a fair comparison in the context of the
present report.
The results in Ref.\ \onlinecite{ytreberg-seps} were obtained
for very short paths, large perturbations of the shoot point, and
a conformational sampling barrier height of 14.0 $k_BT$.
For consistency with the present studies,
SEPS results were generated with
{\it no} perturbation of the shoot point,
much longer paths, and for a range
of conformational sampling barrier heights.
Both TI and SEPS simulations utilized Brownian
dynamics to propagate the system. For SEPS, paths were generated
as described in the present report (but with no velocity), and accepted
with the probability given in Eq.\ \eqref{eq-Pacc2}.

\begin{table}
\begin{tabular}{lccc}
    \hline \hline
    Barrier ($k_BT$) & SEPS long & SEPS short & TI \\
    \hline
    1.0 &  60 000      & 200 000    & 15 300 \\
    \hline
    2.0 &  120 000     & 500 000    & 35 700 \\
    \hline
    4.0 &  400 000     & 1 000 000  & 204 000 \\
    \hline
    6.0 &  1 400 000   & 1 400 000  & 1 020 000 \\
    \hline
    8.0 &  8 000 000   & 1 600 000  & 5 100 000 \\
    \hline
    10.0 & 40 000 000  & 2 400 000  & 20 400 000 \\
    \hline
    12.0 & 80 000 000  & 4 000 000  & 76 500 000 \\
    \hline
    14.0 & 200 000 000 & 10 000 000 & 204 000 000 \\
    \hline \hline
    \end{tabular}
    \caption{
	Number of dynamics steps necessary to be within 0.5 $k_BT$
	of the analytical result for $\Delta F$ with a 0.5 $k_BT$
	or less standard deviation for the
	two-dimensional model in \cite{ytreberg-seps}.
	The first column is the barrier height of the potential
	energy surface in $k_BT$ units. The second and third columns are
	the number of dynamics steps using
	SEPS with, respectively, 200 work values and 20 000 work values.
	The fourth column is the number of dynamics steps using TI with
	using 51 equally spaced values of $\lambda$.
	For both TI and SEPS, half of the generated data were thrown
	away for equilibration.
    \label{tab-2d}
    }
\end{table}

Results for the two-dimensional model using SEPS and TI are shown
in Table \ref{tab-2d}. The free energy change is for switching
between a single-well potential and a double-well potential with
a conformational barrier height in $k_B T$ units given in the first column.
The next three columns give the number of dynamics steps needed
for the \df\ estimate to be within 0.5 $k_BT$ of the correct value
with 0.5 $k_BT$ or smaller standard deviation (estimated over
at least 100 trials): the second and third columns are for SEPS
where either 200 (long trajectories) or 20 000 (short trajectories) 
work values were generated with 50\% of the work values discarded
for equilibration, and the fourth column is TI using 51 evenly spaced
values of $\lambda$ with 50\% of the data at each value of $\lambda$
discarded for equilibration.

Table \ref{tab-2d} clearly shows that, for very low
conformational barrier height,
TI is much more efficient than SEPS, and that the most
efficient SEPS is obtained using longer paths and thus fewer work values.
For increasing conformational barrier heights,
SEPS using long paths and TI become
comparable, while SEPS using short paths becomes the most efficient.
For the largest conformational barrier
height tested in this study (14.0 $k_BT$), SEPS using short paths is
at least 20 times more efficient than either TI or SEPS using long paths.

Since the results for growing and charging an ion in solvent showed
that TI was more efficient than SEPS, we suggest that the free
energy landscapes for the molecular systems used in this study have
rather modest conformational sampling barriers
\cite{elber-barriers,zuckerman-barriers}.

\section{Conclusions}
We have carefully studied several computational
free energy difference (\df) methods,
comparing efficiency and precision. The test cases used for the
comparison were relative solvation energy calculations involving either a
large change in the Lennard-Jones size
or in the charge of a particle in explicit solvent.
Specifically, we compared:
adaptive integration (AIM) \cite{swendsen-aim};
thermodynamic integration (TI) \cite{kirkwood};
path sampling of non-equilibrium work values using both
a Jarzynski uni-directional formalism (SEPS) \cite{ytreberg-seps},
and a Bennett-like bi-directional formalism (BSEPS);
Jarzynski (Jarz) \cite{jarzynski} and Bennett (BJarz)
\cite{crooks-pre,shirts-prl} averaging of non-equilibrium work values;
equilibrium Bennett (Benn) \cite{bennett}; and
free energy perturbation (forward, FEPF and reverse FEPR) \cite{zwanzig}.

AIM \cite{swendsen-aim} was found to provide very high quality,
precise estimates, given long simulation times ($10^6$
total dynamics steps in this study), and also allowed
very rapid estimation of \df.
In addition, AIM provided smooth free energy profiles (and thus smooth
potential of mean force curves) as compared to TI;
see Figs.\ \ref{fig-grow_profile} and \ref{fig-charge_profile}.
Clearly, AIM was the best all-around choice for the systems
studied here.

BJarz \cite{crooks-pre} was also found to perform very well, with
long-simulation results that were second only to AIM. However, it should
be noted that the data shown in this study are for the
most efficient path lengths only.
To determine the optimal path length, many simulations were performed,
adding to the overall cost of the method.
Also, our results showed that using bi-directional data
(BJarz) produced considerably more precise
results than using uni-directional data (Jarz).

The SEPS method is shown to provide accurate free energy estimates
from ``fast-growth''
uni-directional non-equilibrium work values. Specifically,
in cases where the standard deviation of the work values is much greater
than $k_BT$ ($\sigma_W \gg k_BT$),
the SEPS method can effectively shift the work values
to allow for more accurate \df\ estimation than is possible using ordinary
Jarzynski averaging.
Interestingly, using bi-directional data (BSEPS) did not increase
the precision of the \df\ estimate, and perhaps made it somewhat
worse.

We also find, in agreement with previous studies \cite{hummer,crooks-pre},
that the greatest efficiency for the Jarz approach is when
$\sigma_W \approx 1 \; k_B T$. For the first time, we also show
that SEPS is also most efficient when $\sigma_W \approx 1 \;k_BT$,
for the systems studies in this report.

We have also suggested an explanation---with potentially quite
interesting consequences---for the decreased effectiveness of SEPS
in molecular systems. By re-examining the two-dimensional model used
in our first SEPS paper \cite{ytreberg-seps}, we find that
SEPS can indeed be much more more efficient than TI, but
{\it only when the conformational sampling barrier is very high}
($\gg k_BT$). This suggests that the configurational
sampling barriers encountered
in the molecular systems studied here are fairly modest, counter
to our own expectations.
A key question is thus raised: How high are
conformational sampling barriers encountered
in free energy calculations of ``practical interest?'' See also
Refs.\ \onlinecite{elber-barriers,zuckerman-barriers}.

We remind the reader that the results of this study are valid only for
the types of \df\ calculations we considered---namely, growing and charging
a Lennard-Jones particle in explicit solvent.
When large conformational changes are important, such as for binding
affinities, the results could be significantly different---particularly
if large conformational sampling barriers are present.

\section*{Acknowledgments}
The authors would like to thank Ron White
and Hagai Meirovitch for valuable discussions,
and also Manuel Ath\`{e}nes and Gilles Adjanor
for helpful comments regarding the manuscript.
Funding for this research was provided by
the Dept.\ of Computational Biology and
the Dept.\ of Environmental and Occupational Health
at the University of Pittsburgh,
and the National Institutes of Health
(Grants T32 ES007318 and F32 GM073517).

\bibliography{/home/marty/res/tex/my}

\end{document}